\documentstyle[aps,twocolumn]{revtex}

\newcommand{\beq}{\begin{equation}}
\newcommand{\eeq}{\end{equation}} 
\newcommand{\beqa}{\begin{eqnarray}}
\newcommand{\eeqa}{\end{eqnarray}}

\def\half{\frac{1}{2}}
\def\opone{\leavevmode\hbox{\small1\kern-3.8pt\normalsize1}}
 
\begin{document}

\title{Open System Dynamics with Non-Markovian Quantum Trajectories}
\author
{W. T.  Strunz$^{1,\dagger}$, L. Di\'osi$^2$, and N. Gisin$^3$\\
\protect\small\em $^1$Fachbereich Physik, Universit\"at GH Essen, 
45117 Essen, Germany\\
\protect\small\em $^2$Research Institute for Particle and Nuclear 
Physics, 1525 Budapest 114, POB 49, Hungary \\
\protect\small\em $^3$Group of Applied Physics, University of Geneva, 
1211 Geneva 4, Switzerland}
\date{\today}

\maketitle

\abstract{
A non-Markovian stochastic Schr\"odinger equation for
a quantum system coupled to an environment
of harmonic oscillators is presented. 
Its solutions, when averaged over the noise, reproduce the standard 
reduced density matrix without any approximation.
We illustrate the power of this approach with several examples, 
including exponentially decaying
memory correlations and extreme non-Markovian periodic cases, where
the `environment' consists of only a single oscillator. The latter 
case shows the decay and revival of a `Schr\"odinger cat' state.
For strong coupling to a dissipative environment with memory, 
the asymptotic state can be reached in a finite time.
Our description of open systems is compatible with different 
positions of the `Heisenberg cut' between system and environment.}



The dynamics of open quantum systems is a very timely problem, both 
to address fundamental questions (quantum decoherence, measurement 
problem) as well as to tackle the more practical problems of 
engineering the quantum devices necessary for the emerging fields 
of nanotechnology and quantum computing. So far, the true dynamics of
open systems has almost always been simplified by the Markov 
approximation: environmental 
correlation times are assumed negligibly short compared to the system's
characteristic time scale. 

For the numerical solution of Markovian
open systems, described by a master equation of Lindblad form
\beq\label{Lindblad}
\frac{d}{dt}\rho_t= -i[H,\rho_t] +\frac{1}{2}
\sum_m\left([L_m\rho_t, L_m^{\dagger}] + [L_m,\rho_t L_m^{\dagger}]
\right)
\eeq
(where $\rho_t$ denotes the density matrix, $H$ the system's Hamiltonian 
and the operators $L_m$ describe the effect of the environment in the 
Markov approximation), a breakthrough was achieved through the 
discovery of stochastic unravellings \cite{jumps,QSD}.
These are stochastic Schr\"odinger equations for states 
$\psi_t(z)$, driven by a certain classical noise $z_t$ with
distribution functional $P(z)$. Crucially, the ensemble mean 
$M[\ldots]$ over the noise
recovers the density operator,
\beq\label{rhopsi}
\rho_t = M\Bigl[|\psi_t(z)\rangle\langle\psi_t(z)|\Bigr].
\eeq
Hence, the solution
of eq.(\ref{Lindblad}) is reduced from a problem in the matrix
space of $\rho$ to a much simpler Monte Carlo simulation of
quantum trajectories $\psi_t(z)$ in the state space.

For the Markov master eq.(\ref{Lindblad}), several such unravellings 
are known.  Some involve jumps at random times \cite{jumps}, 
others have continuous, diffusive solutions \cite{QSD}.  
They have been used extensively over recent years, 
as they provide useful insight into the dynamics of continuously 
monitored (individual) quantum processes \cite{insight}, or into
the mechanism of decoherence \cite{decoherence}.
In addition, they provide an efficient tool for the numerical solution 
of the master eq.(\ref{Lindblad}).
It is thus desirable to extend the powerful concept of stochastic
unravellings to the more general case of non-Markovian evolution. 

The simplest unravellings are linear stochastic Schr\"odinger equations. 
In the Markov case (\ref{Lindblad}), for a single $L$, the linear equation
\beq\label{MLQSD}
\frac{d}{dt}\psi_t = -iH\psi_t 
+ L\psi_t\circ z_t
- \frac{1}{2} L^\dagger L \psi_t,
\label{dpsiMarkov}
\eeq
provides such an unravelling,
where, $z_t$ is a complex-valued Wiener process of zero mean and
correlations
$M[z_t^*z_s]=\delta(t-s)$, $M[z_t z_s]=0$,
and where $\circ$ denotes the Stratonovich product \cite{HasEza80}.

However, eq.(\ref{MLQSD}) is of limited value, since
the norm $\Vert\psi_t(z)\Vert$ of its solutions tends to 0 with 
probability 1 and to infinity with probability 0, such that
the mean square norm is constant. 
To be really useful, one should find unravellings in terms of the 
normalized states 
\beq\label{normpsi}
\tilde\psi_t(z)=\frac{\psi_t(z)}{\Vert\psi_t(z)\Vert},
\eeq
which  requires a
redefinition of the distribution of the noise
$P(z) \rightarrow \tilde P_t(z)\equiv\Vert\psi_t(z)\Vert^2 P(z)$
\cite{GirsTrans}
so that eq.(\ref{rhopsi}) remains valid for the normalized solutions:
\beq\label{rhonormpsi}
\rho_t = \tilde M_t\Bigl[|\tilde\psi_t(z)\rangle\langle\tilde\psi_t(z)|
\Bigr].
\eeq
Now (\ref{rhonormpsi}) can be interpreted as
an unravelling of the mixed state $\rho_t$ into an ensemble of pure 
states. 
For the Markov unravelling (\ref{MLQSD}), the normalized states 
$\tilde\psi_t$ satisfy the non-linear Quantum State Diffusion (QSD) 
equation \cite{QSD}:
\beqa\label{MQSD}
\frac{d}{dt}
\tilde\psi_t &=& -iH\tilde\psi_t 
 + (L-\langle L\rangle_t)\tilde\psi_t\circ 
(z_t + \langle L^\dagger\rangle_t) 
\\ \nonumber
& &  
-\frac{1}{2} (L^\dagger L-\langle L^\dagger L\rangle_t)\tilde\psi_t,
\eeqa
where $\langle L\rangle_t\equiv\langle\tilde\psi_t|L
|\tilde\psi_t\rangle$.
Contrary to eq.(\ref{MLQSD}), eq.(\ref{MQSD}) provides an efficient 
Monte-Carlo algorithm for the numerical solution
of (\ref{Lindblad}) \cite{jumps,QSD}.

In this Letter we present for the first time a nonlinear non-Markovian
stochastic Schr\"odinger equation that unravels the dynamics of a 
system interacting with an arbitrary `environment' of 
a finite or infinite number of harmonic oscillators, without any
approximation.  In the Markov limit, this unravelling
reduces to QSD (\ref{MQSD}) and will therefore be referred to as 
non-Markovian Quantum State Diffusion. Other authors have treated 
non-Markovian open systems effectively with Markovian unravellings:
either the system has in fact been influenced by a second Markovian
environment in addition to the original non-Markovian one or, 
alternatively, fictious modes have been added to the system 
\cite{nonmarkov}. In our approach, the `system' remains 
unaltered, and the unravelling is genuinely non-Markovian.

Below we summarize the general theory, which will be presented in detail
elsewhere \cite{DiosiGisinStrunz98}, and we present four examples:
First, we consider a `measurement-like' environment. Then, a dissipative 
environment with exponentially decaying environment
correlations is discussed. Remarkably, here the 
asymptotic state can be reached in a finite time.
In the third example we consider an `environment' consisting of only a 
single oscillator. This example is thus periodic, that is extremely 
non-Markovian. 
It shows the decay and revival of a 'Schr\"odinger cat' state.
Finally, the fourth example shows that the description of a 
subsystem in terms of non-Markovian QSD is independent of the 
`Heisenberg cut', that is independent of where precisely the boundary 
between system and environment is set. 

Our starting point is the non-Markovian generalization of the
linear stochastic equation (\ref{MLQSD}), derived in \cite{LNMQSD},
\beq\label{LQSD}
\frac{d}{d t}\psi_t=-iH\psi_t 
+ L\psi_t z_t
- L^\dagger\int_0^t \alpha(t,s)\frac{\delta\psi_t}{\delta z_s}ds ,
\eeq
which unravels the exact reduced dynamics of a system coupled to an
environment of harmonic oscillators.
Here, $z_t$ is colored
complex Gaussian noise of zero mean and correlations
\beq\label{correlation}
M\left[z_t^*z_s\right]=\alpha(t,s),\;\;\; M\left[z_t z_s\right]=0.
\eeq
The Hermitian $\alpha(t,s) = \alpha^*(s,t)$ is the environment
correlation function \cite{DiosiGisinStrunz98,LNMQSD}. 
The functional derivative under the memory integral 
in (\ref{LQSD}) indicates 
that the evolution of the state $\psi_t$ at time
$t$ is influenced by
its dependence on the noise $z_s$ at earlier times s. 
In \cite{DiosiGisinStrunz98} we show that it
amounts to applying an operator 
to the state,
\beq\label{strange}
\frac{\delta}{\delta z_s} \psi_t \equiv \hat O (s,t,z) \psi_t,
\eeq
where the explicit expression of $\hat O(s,t,z)$ can be determined 
consistently from eq.(\ref{LQSD}).

Just as in the Markov limit, to be really useful, one has to find 
the corresponding non-linear non-Markovian QSD equation for
the normalized states (\ref{normpsi}). This quite elaborate
derivation can be found in \cite{DiosiGisinStrunz98} and leads
to
\beqa\label{NMQSD}
&&\frac{d}{dt}\tilde\psi_t = 
-iH\tilde\psi_t
+(L-\langle L \rangle_t) \tilde\psi_t \tilde z_t \\ \nonumber
& & - \int_0^t \alpha(t,s)
\left(\Delta L^\dagger \hat O(s,t,\tilde z)
- \langle \Delta L^\dagger\hat O(s,t,\tilde z)
\rangle_t\right)ds\tilde\psi_t,
\eeqa
which is the basic equation of non-Markovian QSD. Here,
$\tilde z_t$ is the shifted noise
$\tilde z_t = z_t + \int_0^t\alpha^*(t,s)\langle L^\dagger\rangle_s ds$,
and for brevity we use $\Delta L^\dagger = 
L^\dagger - \langle L^\dagger\rangle_t$.

Let's turn to concrete examples of non-Markovian QSD
(\ref{NMQSD}). First, we 
consider an environment modeling energy 
measurement: $L=L^\dagger=H$. It is easy to prove
that $\hat O = H$ in (\ref{strange}),
and hence (\ref{NMQSD}) reads
\beqa\label{NMQSDH}
&\frac{d}{dt}&\tilde\psi_t = 
-iH\tilde\psi_t - (H^2-\langle H^2\rangle_t)\tilde\psi_t
\int_0^t\alpha(t,s)ds \\ \nonumber
&+&(H - \langle H\rangle_t)\tilde\psi_t
\left(z_t+\!\int_0^t\!\!\alpha(t,s)^*\langle H\rangle_s ds
+\!\int_0^t\!\!\alpha(t,s)ds\langle H\rangle_t\right).
\eeqa
Notice that indeed, (\ref{NMQSDH}) reduces to the Markov QSD 
equation (\ref{MQSD}) for
$\alpha(t,s)\rightarrow \delta(t-s)$.

If the correlation $\alpha(t,s)$ 
decreases fast enough, the 
asymptotic solution of (\ref{NMQSDH}) is an eigenstate 
$\phi_n$ of $H$,
reached with the expected quantum probability
$|\langle\phi_n|\psi_0\rangle|^2$. Numerical
solutions of (\ref{NMQSDH}) for the 2-dimensional case 
$H=\frac{\omega}{2}\sigma_z$ and exponentially decaying correlation
are shown in Fig.1a (solid lines).
The asymptotic state 
is either the 'up' or the 'down' state. The ensemble mean
$M[\langle\sigma_z\rangle]$ remains constant (dashed line) as 
expected from the analytical solution (dot-dashed line).
Note, however, that if the environment consists of a finite number of 
oscillators,
represented by a quasi-periodic correlation function $\alpha(t,s)$,
such a reduction to an eigenstate will not occur (see our third 
example).

As a second example, we consider a dissipative spin 
with $H=\frac{\omega}{2}\sigma_z$, and
$L=\lambda\sigma_-$. We choose exponentially decaying correlations
$\alpha(t,s)=\frac{\gamma}{2}e^{-\gamma|t-s|-i\Omega(t-s)}$ with an
environmental central frequency $\Omega$ and  memory time $\gamma^{-1}$. 
The non-Markovian QSD equation (\ref{NMQSD}) reads
\cite{DiosiGisinStrunz98}
\beqa\label{NMQSDsigmadamped}
&&\frac{d}{dt}{\tilde\psi}_t=-i\frac{\omega}{2}\sigma_z\tilde\psi_t-\lambda
F(t)(\sigma_+\sigma_--\langle\sigma_+\sigma_-\rangle_t)\tilde\psi_t \\ 
\nonumber
&+&\lambda(\sigma_--\langle\sigma_-\rangle_t)\tilde\psi_t
\left(z_t+\lambda\int_0^t\alpha(t,s)^*\langle\sigma_+\rangle_s ds
+\langle\sigma_+\rangle_t F(t)\right)
\eeqa
with $F(t)$ determined from 
\beq
\frac{d}{dt} F(t)=-\gamma F(t) + i(\omega-\Omega)F(t) + \lambda F(t)^2 
+ \frac{\lambda\gamma}{2}
\label{dotF}
\eeq
and initial condition $F(0)=0$. The equation for $F(t)$ can be solved 
analytically \cite{DiosiGisinStrunz98}.
It is worth mentioning the case of exact resonance, $\omega=\Omega$.
Two regimes should be distinguished. First, 
when $\gamma > 2\lambda^2$ (short memory compared to coupling strength), 
$F(t)$ tends to
$\left(\gamma-\sqrt{\gamma^2-2\gamma\lambda^2}\right)/(2\lambda)$.
Hence, for large $\gamma$, one recovers Markov QSD (\ref{MQSD}).
For longer memory times or stronger coupling, 
$\gamma < 2\lambda^2$, things are very different:
$F(t)$ diverges to infinity when the time $t$ approaches
$t_c=\left(\pi+2\arctan(\gamma/\sqrt{2\lambda^2\gamma-\gamma^2})\right)/
\sqrt{2\lambda^2\gamma-\gamma^2}$. 
All realizations $\psi_t(z)$
reach the down state in a finite time and remain there! 
In Fig.1b we show quantum trajectories
from (\ref{NMQSDsigmadamped}) (solid lines),
their ensemble mean value $M[\langle\sigma_z\rangle_t]$
(dashed line), and the analytical
mean value (dot-dashed), which is almost indistinguishable.
The reduction time in this case is
$\omega t_c = \frac{3}{2}\pi \approx 4.71$. 
This is the first
example of a continuous quantum state diffusion that reaches its 
asymptotic state in a finite time, which
was proven impossible for Markovian diffusions \cite{tails}.


Our third example is a harmonic oscillator coupled to a 
finite or infinite number
of oscillators initially in their ground states. Here,
the non-Markovian QSD eq.(\ref{NMQSD}) takes the same form 
(\ref{NMQSDsigmadamped}), where the
Hamiltonian is $\omega a^\dagger a$ and where $\sigma_-$ ($\sigma_+$) has
to be replaced by the annihilation (creation) operator $a$ ($a^\dagger$). 
The resulting equation preserves coherent states.
More interesting is the case of an initial superposition
of two symmetric coherent states, known as a `Schr\"odinger cat' 
\cite{cat}. 
If the environment correlation $\alpha(t,s)$
decays, so does the `cat'. If, however,
the environment consists of only a finite number of
oscillators, then the `cat' will first decay, due to the localization 
property of QSD,
but since the entire system is quasi-periodic, the `cat' will 
then revive! 
In Fig.2 we show contour plots of the evolution of the $Q$-function 
of such a `cat', in
the extreme case where the environment consists of a single 
oscillator ($\alpha(t,s)=e^{-i\Omega(t-s)}$).
Apart from an overall
spiraling motion due to the `system' Hamiltonian, the `cat' state first 
decays but later revives.
Our non-Markovian QSD equation thus
provides a nice illustration of proposed experiments on reversible
decoherence \cite{Haroche}. 


As a last example we consider a case where the split between 
system and environment
can be shifted naturally between two positions, see Fig.3. 
A spin (Hamiltonian $H_1$) and a distinguished harmonic oscillator 
($H_2$) are linearly coupled ($H_{12}$). Moreover, the spin is coupled
($H_I$) to a heat bath ($H_{env}$) at zero temperature.
We can either consider the quantum state of the 
spin-oscillator system coupled to a heat bath, or the 
quantum state of
the spin coupled to a heat bath and coupled to the distinguished 
oscillator. In the 
first case, we can apply the Markov QSD description, ie a family of 
spin-oscillator states $\psi_t(\xi)$ indexed by a complex Wiener 
process $\xi_t$.
In the second case, using non-Markovian QSD, we have a family of spin 
states
$\phi_t(\xi,z)$ indexed by the same $\xi_t$ and also by the 
noise $z_t$, due to the distinguished oscillator `environment'
with correlation
$M[z_t^*z_s] = e^{-i\omega_2 (t-s)}$.

Let us study a shift of the `Heisenberg cut': compare
the states $\phi_t(\xi,z)$ of the spin averaged over the noise $z_t$ 
with the mixed state
obtained by tracing out (Tr$_2$) the oscillator from the spin-oscillator 
states $\psi_t(\xi)$. We prove in \cite{DiosiGisinStrunz98} that the 
states corresponding to both descriptions are equal:
\beq
M_z[|\phi_t(\xi,z)\rangle\langle\phi_t(\xi,z)|] = 
\mbox{Tr}_2(|\psi_t(\xi)\rangle\langle\psi_t(\xi)|).
\label{equ}
\eeq
This illustrates the general fact that non-Markovian QSD attributes 
stochastic pure states to a system in a way which depends on the position 
of the Heisenberg cut, but which is
consistent for all possible choices of the cut. 


In conclusion, we present the first non-Markovian unravelling 
of the dynamics of a quantum 
system coupled to an environment of harmonic oscillators,
which can thus be simulated by classical complex noise.
In the Markov limit, standard Quantum State Diffusion is recovered. 
We emphasize that non-Markovian QSD (\ref{NMQSD}) reproduces the
true evolution of the system taking into account the exact unitary
dynamics of system and environment \cite{DiosiGisinStrunz98,LNMQSD}.
The power of this new approach to open quantum systems
is illustrated with four examples. For measurement-like interactions, 
reduction to eigenstates takes place
whenever the environment correlation function decreases fast enough. 
For dissipative
interaction with a heat bath at zero temperature, the ground 
state may be reached in a finite time.
The third example is an application to the most extreme 
non-Markovian case:
two linearly coupled oscillators, one of them playing the role of the 
`environment'.
We see the decay and revival of a `Schr\"odinger cat' state. Finally,
the last example illustrates that unravellings corresponding to 
different positions of the `Heisenberg cut' between system and
environment are mutually compatible. 
Most of these features are entirely new
and have no counterpart in any Markov unravelling. Hence, 
non-Markovian unravellings represent a promising route to open 
systems, 
as for instance to quantum Brownian motion \cite{DiosiGisinStrunz98}. 
Moreover, our approach represents a
new efficient tool for the numerical simulation of quantum 
devices, whenever non-Markovian effects are relevant
\cite{nonmarkov,devices}.

We thank IC Percival for helpful comments and the University of Geneva 
where part of the work was done.
WTS would like to thank
the Deutsche Forschungsgemeinschaft for support through the SFB 237
"Unordnung und gro{\ss}e Fluktuationen".
LD is supported by the Hungarian Scientific Research Fund through grant
T016047.
NG thanks the Swiss National Science Foundation.

\newpage

\section*{Figure Captions}

\vspace{.3cm}
{FIG. 1. {\small{Non-Markovian quantum trajectories (solid lines) for 
a spin $\half$ system $H=\frac{\omega}{2}\sigma_z$ with an 
exponentially decaying environment correlation 
$\alpha(t,s) = \frac{\gamma}{2}\exp\left(-\gamma|t-s|-i\Omega(t-s)\right)$,
where we choose $\gamma = \omega$. The ensemble mean value 
over $10000$ runs (dashed line) is in very good agreement with
the analytical result (dot-dashed line).
We show
(a) a measurement-like interaction $L=\lambda \sigma_z$ with 
$\Omega = 0$, and (b)
a dissipative interaction $L=\lambda \sigma_-$ on resonance
$\Omega = \omega$, where each trajectory reaches the ground state
in a finite time $\omega t_c = \frac{3}{2}\pi \approx 4.71.$
In both cases we choose a coupling strength $\lambda^2 = \omega$ 
and an initial state
$|\psi_0\rangle = 3|\uparrow\rangle + 2|\downarrow\rangle$.
}}}\\

\vspace{.3cm}
{FIG. 2. {\small{Reversible decoherence of an initial symmetric
`Schr\"odinger cat' state 
$|\psi_0\rangle = |\alpha\rangle + |-\alpha\rangle$ with $\alpha=2$.
The contour plots show the $Q$-function of 
a non-Markovian quantum trajectory of a 
harmonic oscillator ($\omega$) `system', coupled to just a single 
`environment'
oscillator ($\Omega = 0.5\omega$). The coupling strength between the
two oscillators is $0.1\omega$,
and the time step between two successive plots is $0.47/\omega$.
}}}\\

\vspace{.3cm}
{FIG. 3. {\small{
`Spin -  single oscillator - heat bath' system. First, we
consider the `spin -  single oscillator' as the `system' with
state $\psi_t(\xi)$, coupled to the heat bath with
noise $\xi_t$.  Alternatively, we can consider the
`spin' as the `system' $\phi_t(\xi,z)$, coupled to the 
`single oscillator $+$ heat bath' environment 
(noises $(\xi_t,z_t)$). In non-Markovian QSD, both
descriptions are possible and lead to the same reduced spin state.
}}}\\

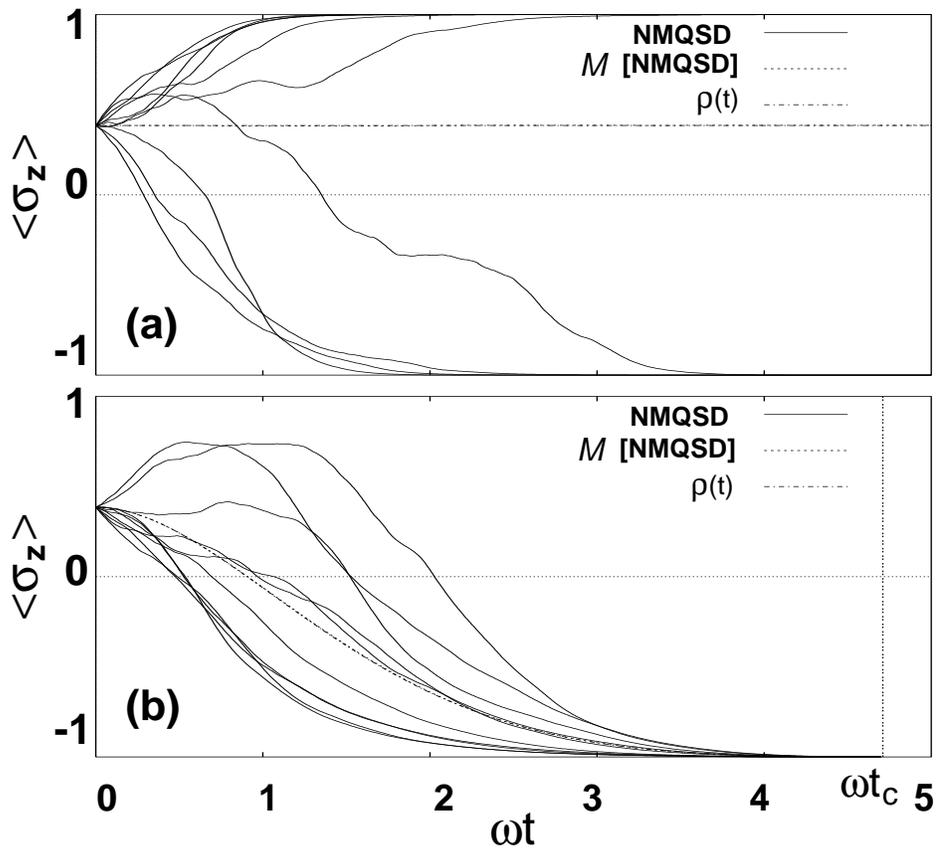
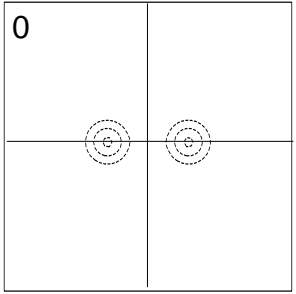
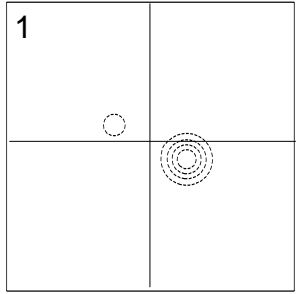
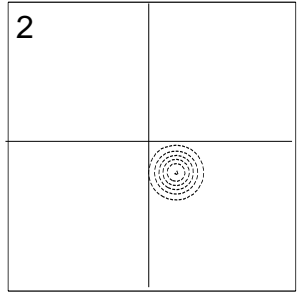
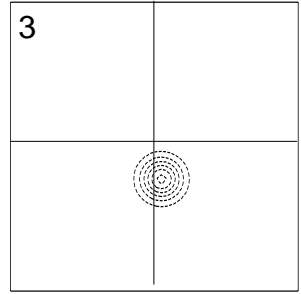
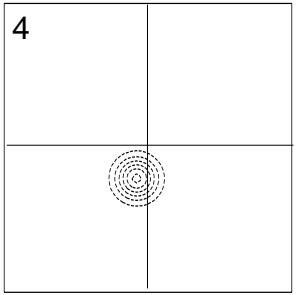
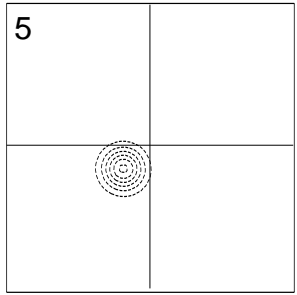
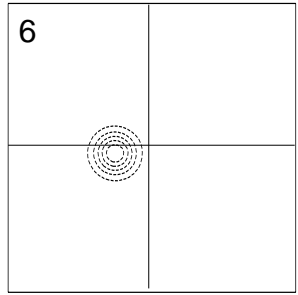
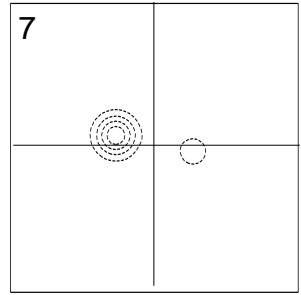
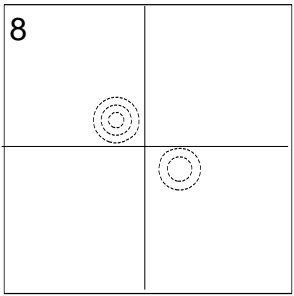
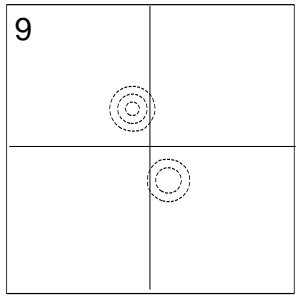
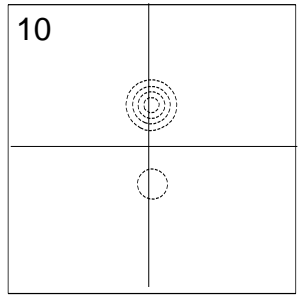
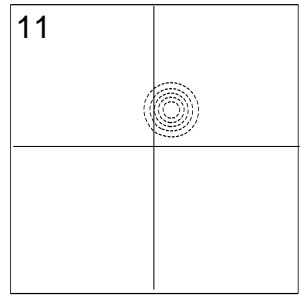
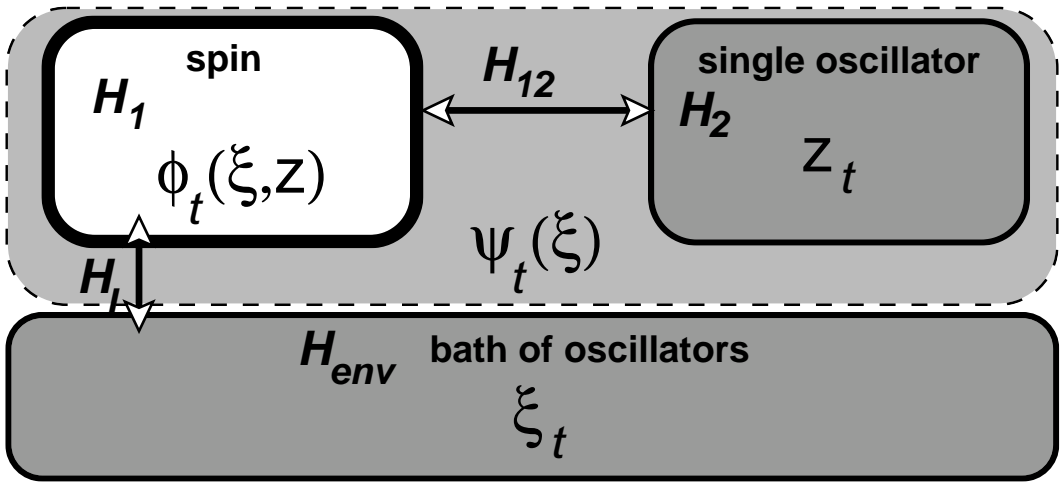
\end{document}